**Re-acceleration of Energetic Ions via Small-Scale Reconnection in Magnetic Fusion Plasmas**

Cong ZHANG[1,#], Shaodong SONG[1,#], Di LUO[1], Kai HUANG[2], Linge ZANG[3], Huibo TANG[2], Yanchao LI[1], Yihang ZHAO[1], Ao WANG[1], Hanqing WANG[1], Zhenxing WANG[1], Lei HAN[1], Xuxu ZHANG[1], Jia LI[1], Dong GUO[1], Yunfeng LIANG[1], Minsheng LIU[1], Yuejiang SHI[1,*]

[1]ENN Science and Technology Development Co., Ltd., Langfang 065001, People's Republic of China

[2] Harbin Institute of Technology, Harbin 150001, People's Republic of China

[3]Southwestern Institute of Physics, Chengdu 610041, People's Republic of China

#ZHANG and SONG contributed equally to this work.
*Corresponding author: yjshi@ipp.ac.cn

**Abstract**: We report the first observation on the EXL-50U spherical torus that energetic particles injected by neutral beam injection (NBI) can be stably accelerated to significantly higher energies - reaching up to 2.5 times the injection energy, occurring without significant large-scale magnetohydrodynamic (MHD) bursts. Simulations based on EXL-50U parameters indicate that small-scale magnetic reconnection, mediated by multiple magnetic islands, fails to accelerate bulk thermal ions but efficiently energizes seed fast ions. Unlike global MHD events, such small-scale reconnection is ubiquitous in magnetic confinement devices and does not degrade core confinement. This mechanism offers a novel and potentially universal channel for auxiliary ion heating in future fusion reactors.

Efficient ion heating is a prerequisite for achieving fusion ignition and sustained burning.In magnetic confinement devices, the external injection of high-power neutral beam injection (NBI) or ion cyclotron resonant frequency (ICRF) waves is generally regarded as the most effective means to elevate the plasma ion temperature to ignition conditions. However, these external heating systems are typically complex, expensive, and occupy valuable port space. Upon entering the burning phase, fusion $\alpha$ particles will serve as the primary heating source for the plasma. Because the energy of these $\alpha$ particles is significantly higher than the electron temperature of the background plasma, if energy transfer occurs solely through collisions, their energy is predominantly deposited into the electron channel. To address this, Nathaniel J. Fisch and his group proposed the $\alpha$-channeling mechanism from 1990s [1-4], which, in principle, allows the energy of $\alpha$ particles to be transferred to background ions via wave-particle interactions. On the other hand, $\alpha$-channeling imposes stringent requirements on the background plasma parameters, profile distributions, and wave fields, making it exceptionally challenging to satisfy the necessary conditions in practical magnetic confinement fusion plasmas. Consequently, for magnetic confinement fusion, identifying a simple and feasible heating method that can substitute or partially substitute external heating and $\alpha$-channeling is of profound importance for accelerating the practical realization of fusion power.

Accelerating ions via magnetic reconnection is a ubiquitous and dominant mechanism in astrophysical plasmas. In several magnetic confinement devices [5-9], plasma startup and initial heating achieved through dedicated merging coils rely on this reconnection mechanism. However, merging coils are typically employed only during the plasma startup phase; significant uncertainty remains as to whether they can be utilized to stabilize and sustain heating during the current flat-top phase. In other devices, such as MAST and VEST [10,11], ion acceleration has been observed during internal reconnection events (IREs) in the flat-top phase of the discharge. This ion heating is attributed to the magnetic reconnection acceleration mechanism triggered by the IRE. Nevertheless, following the onset of large-scale magnetohydrodynamic (MHD) events like IREs, plasma parameters and overall performance degrade significantly, meaning that the ion heating induced by IREs is merely bursty, intermittent, and unsustainable. In this Letter, we report the first observation of stable and sustained high-energy ion acceleration on the EXL-50U [12,13] device, achieved in the absence of IREs or other large-scale MHD instabilities.

We have reproduced the MAST experimental results [10] regarding ion acceleration during IREs in Ohmic heating plasmas on the EXL-50U device. In shot #16478, a typical IRE is identified by spikes in $I_p$ and loop voltage($V_{\mathrm{loop}}$) [Fig.1]. Multi-channel neutral particle analyzer (NPA) [14,15] measurements reveal a high-energy ion tail reaching ~40 keV. Magnetic probe signals confirm temporal correlation between MHD activity and ion acceleration. However, as shown in Fig. 1, the electron temperature $T_e$ drops sharply after the IRE. Large-scale magnetic reconnection events like IREs, which accelerate ions, come at the expense of a substantial degradation in overall plasma confinement performance.

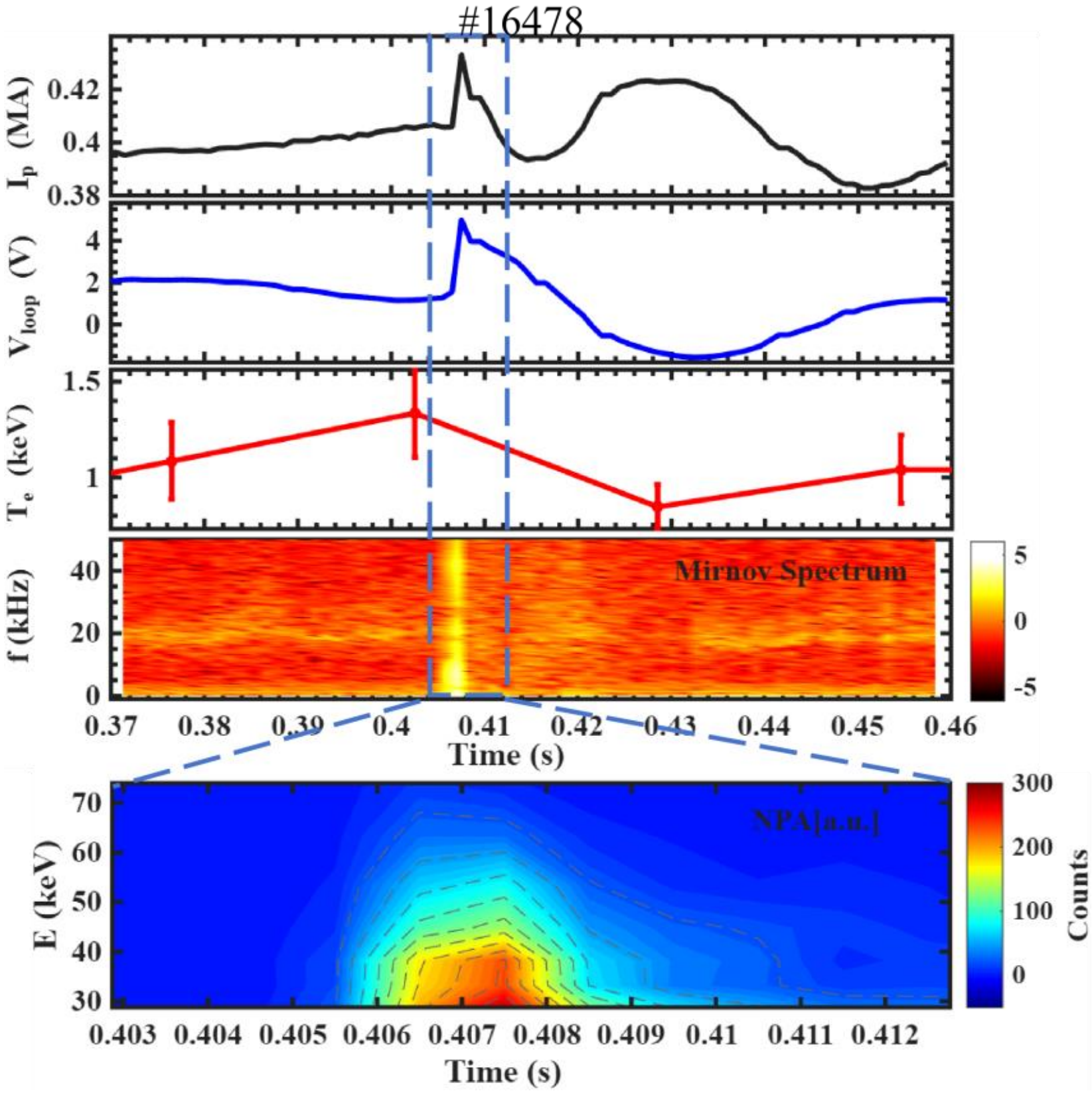

**FIG. 1.** Experimental evidence of ion acceleration during a disruptive Internal Reconnection Event (IRE) in an Ohmic discharge (#16478). From top to bottom: $I_p$, $V_{\mathrm{loop}}$, $T_e$, Mirnov coil spectrogram showing MHD activity, and the time evolution of the energetic ion energy spectrum measured by the NPA. The dashed lines in the bottom panel indicate the rapid energization of ions up to 40 keV correlated with the $I_p$ spike and $T_e$ collapse.

The key finding of this work is that pre-existing fast ions enable efficient acceleration even during small-scale reconnection along with weak MHD activity. We did experiments under different beam injection energies. In these experiments, we have observed a population of high-energy particles whose energies significantly exceed the neutral beam injection energy (see Fig. 2). A more striking phenomenon was observed when electron cyclotron heating (ECH) was performed in these NBI discharges (shot #16534). Comparing the NBI-only case (shot #16531, $E_{\mathrm{inj}}$ =40 keV with the NBI+ECH case (shot #16534), we observe a dramatic extension of the ion energy tail (see Fig. 2). The tail energy can reach about 53 keV when using NBI alone, while the energy can reach 102 keV when using NBI+ECH - which is 2.5 times higher than the injection energy.

In these NBI or NBI+ECH experiments, no IREs were observed. Only weak MHD activity was present, and energetic particles were stably accelerated during the discharge (as shown in fig.3). Mirnov spectrograms reveal only a persistent $m/n$=3/1 mode near 20 kHz, whose magnetic fluctuation amplitude is lower than the intense bursts observed during the disruptive IRE in shot #16478. The electron temperature also maintained a steady upward trend during NBI and ECH.

Based on the above experimental phenomena, we speculate that small-scale magnetic reconnection can accelerate the existing background high-energy particles to higher energy, although it cannot accelerate the low-energy thermal ions. We propose that EC power deposition modifies the local safety factor $q$ profile and magnetic shear, thereby affecting the structure of magnetic reconnection and further accelerating high-energy particles.

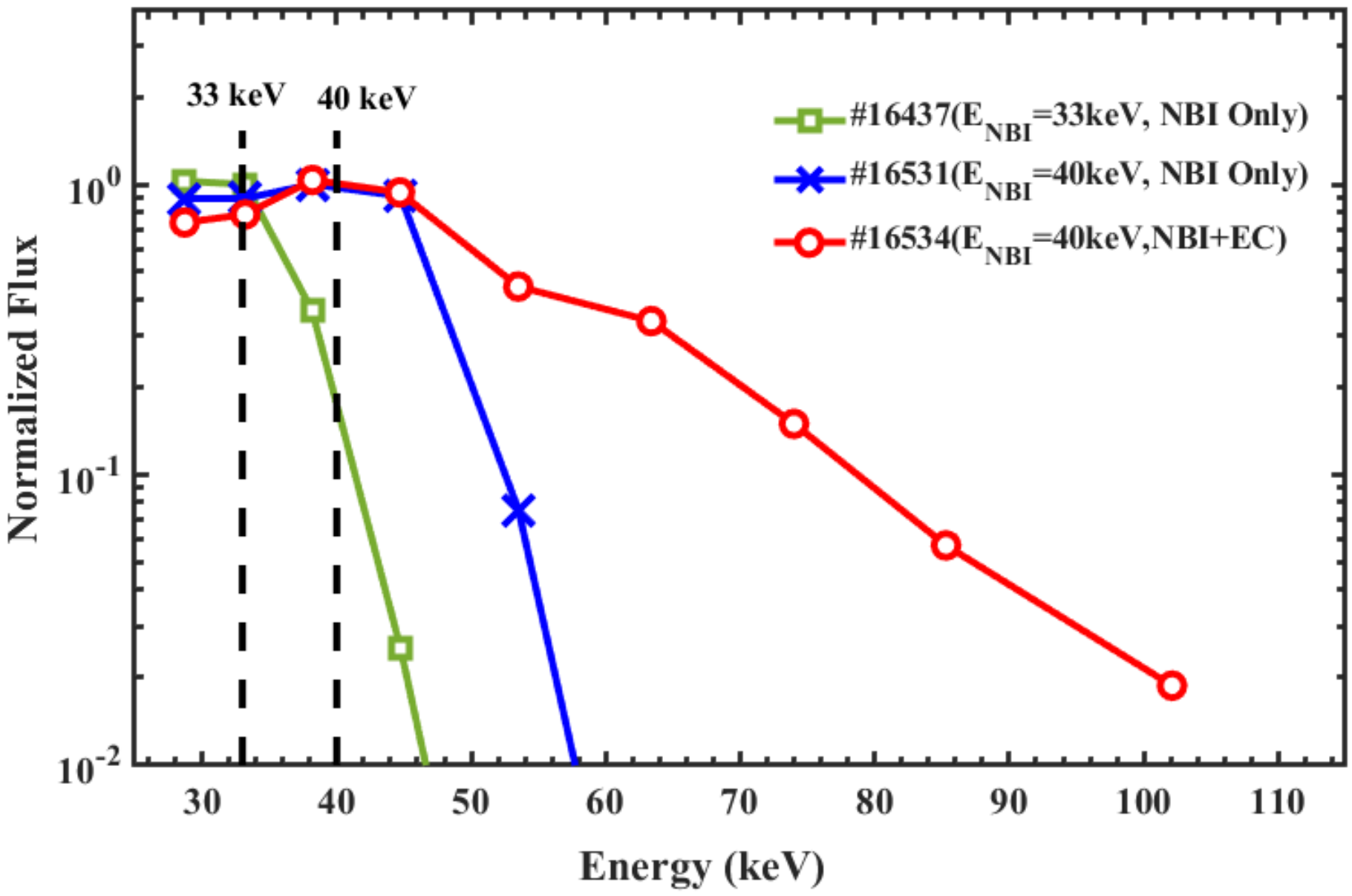

**FIG. 2.** Re-acceleration and active control of NBI-seeded ions during small-scale reconnection. NPA ion energy spectra for NBI-only discharges at different injection energies: Shot #16437 ($E_{\mathrm{NBI}}$ =33 keV, green line with squares), Shot #16531 ($E_{\mathrm{NBI}}$ =40 keV, blue line with cross marker) and the NBI+ECH synergistic discharge (Shot #16534, red line with circles). The application of EC waves extends the high-energy tail up to 102 keV, representing a 2.5 times increase relative to the injected beam energy. All spectra are normalized to the count rate at the NBI injection energy.

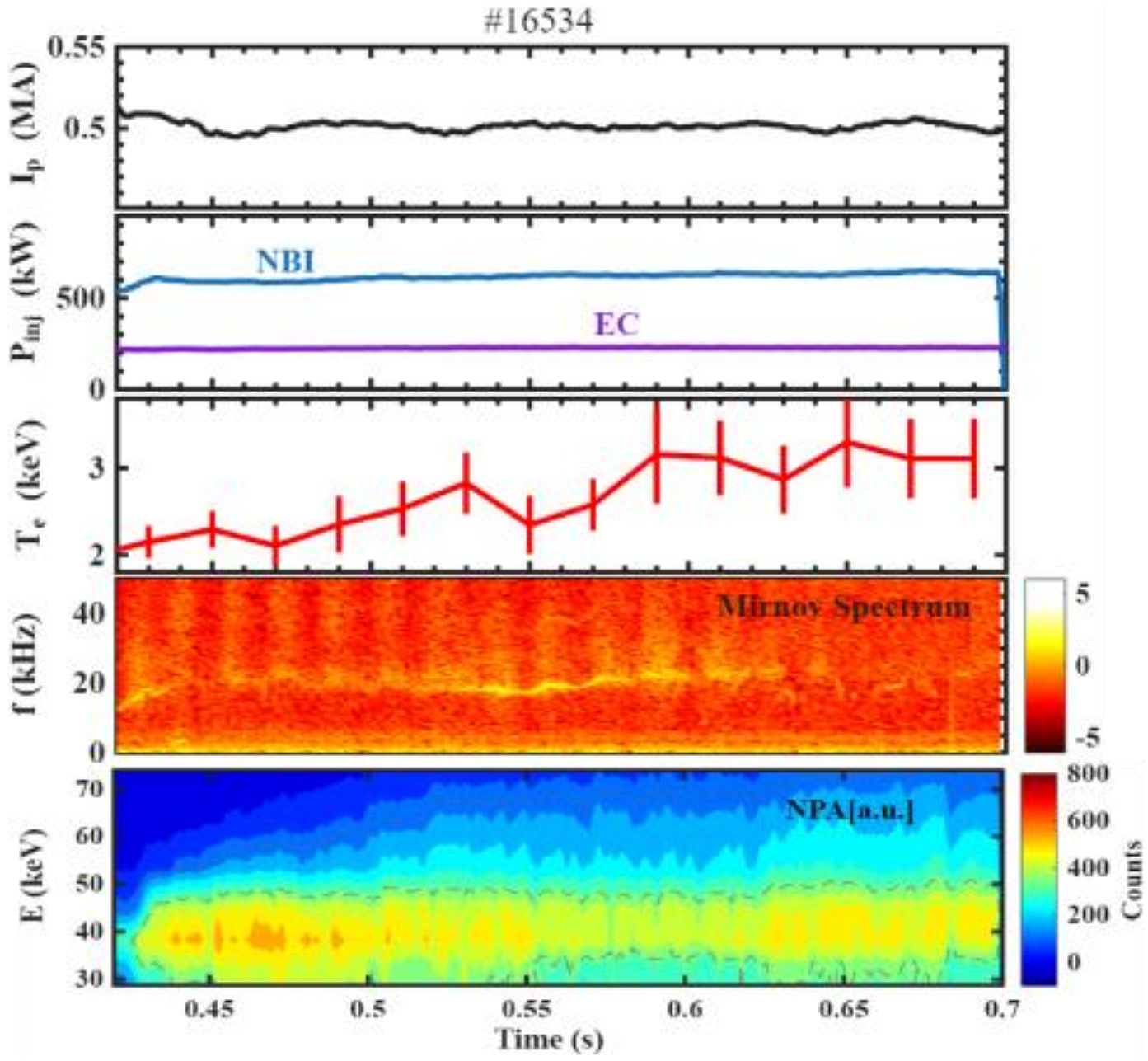

**FIG. 3.** Time evolution of main plasma parameters for #16534 in EXL-50U. From top to bottom: $I_p$, $V_{\text{loop}}$, $T_e$, Mirnov spectrogram, and the time evolution of the energetic ion energy spectrum measured by the NPA. The discharge illustrates the plasma response during small-scale magnetic reconnection events.

To elucidate the physical mechanisms underlying the observed acceleration hierarchy, we performed kinetic simulations tailored to EXL-50U parameters ($B_t$=0.7 T, $n_e$=5.7 $^{18}$m$^{-3}$) using the open-source particle-in-cell code VPIC [16-17].  By varying the spatial scale of the reconnection zone and the current sheet thickness, we modeled the ion acceleration process during magnetic reconnection in the EXL-50U. The influence of the reconnection zone size $L_Z$, the initial current sheet thickness $\delta$, and the presence of pre-existing energetic ions was investigated.

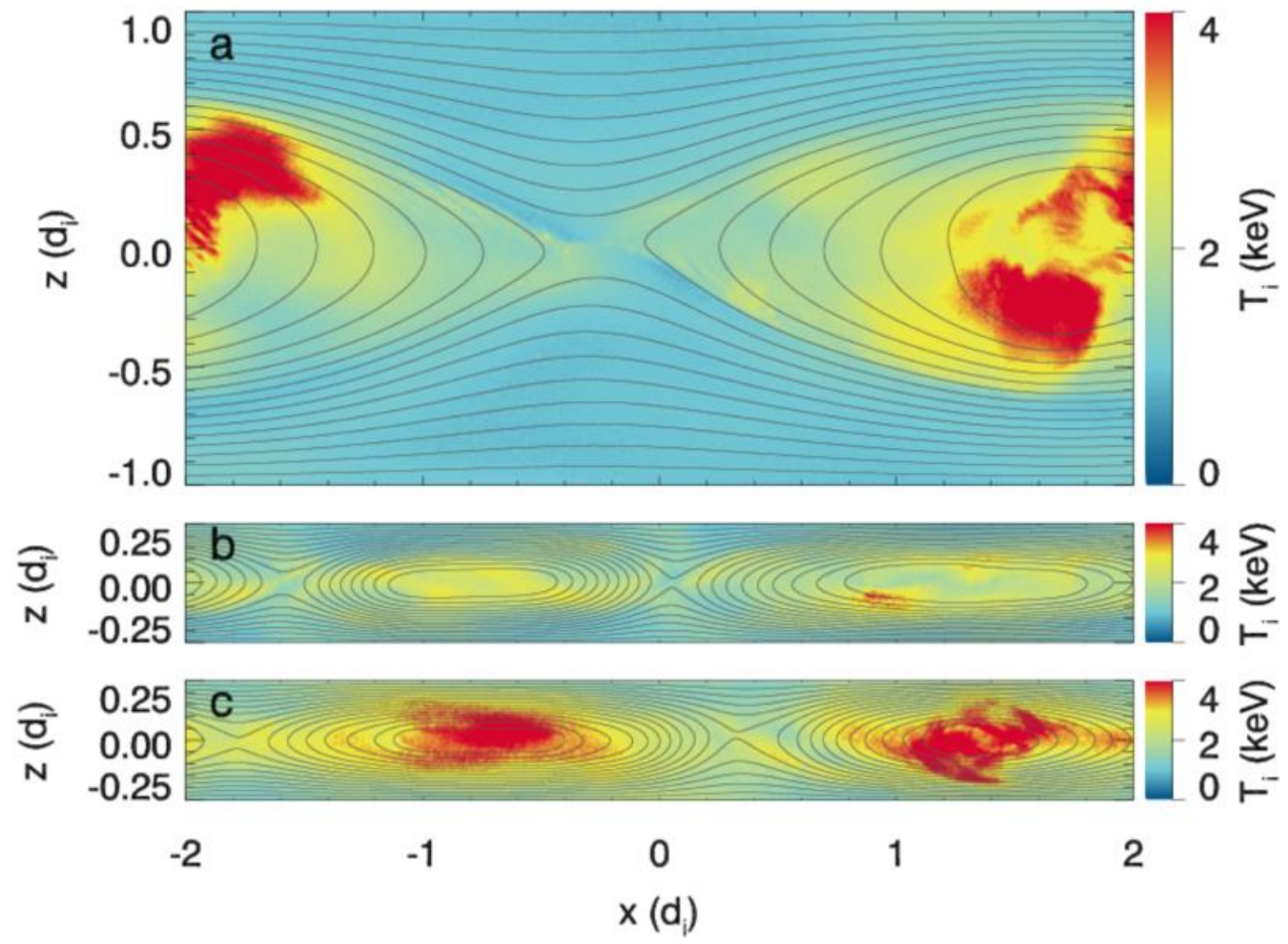

、

**FIG. 4.** Magnetic islands formation during large and small-scale magnetic reconnection in

(a) case 1, (b) case 2, and (c) case 3, the colored contours show the ion temperature, and the in-plane curves represent the magnetic field lines.

Fig. 4(*a*) illustrates the formation of a magnetic island with a width reaching $1d_i$ (approximately 10 cm) during the large-scale reconnection, indicating that large scale reconnection can lead to strong ion acceleration while severely distorting the magnetic field configuration – consistent with the IRE observed in shot #16478. In Fig. 4(*b*) and 4(*c*), $L_Z$ was reduced to $0.5d_i$ to prevent the development of large scale reconnection, resulting in a regime characterized by weak MHD activity. The width of the magnetic island is reduced to less than $0.3d_i$ (corresponding to around 3 cm), and such kind of small scale reconnection will not significantly affect macroscopic magnetic confinement. In Fig. 4(*b*), we added 1% pre-existing energetic ions (40 keV), corresponding to the NBI-injected experimental condition, the 40keV high-energy component is significantly heated, producing energetic ions with energies up to approximately 60keV (red dotted line in Fig. 5). we can conclude that small scale reconnection can further energize the pre-existing high-energy ions, leading to the increase of the energetic ion flux while keeping the magnetic confinement, consistent with the results observed in shot #16531 in Fig. 2. In Fig. 4(*c*), we further reduce the initial thickness of the current sheet based on case in Fig. 4(*b*) to simulate the localized effects of ECH, which is known to modify the local magnetic shear. The energization of the pre-existing high-energy ions is much stronger (blue dotted line in Fig. 5), suggesting that the ECH-driven current sheet thinning can further enhance the production of energetic ions via small scale reconnection, consistent with the results observed in shot #16534 in Fig.2.

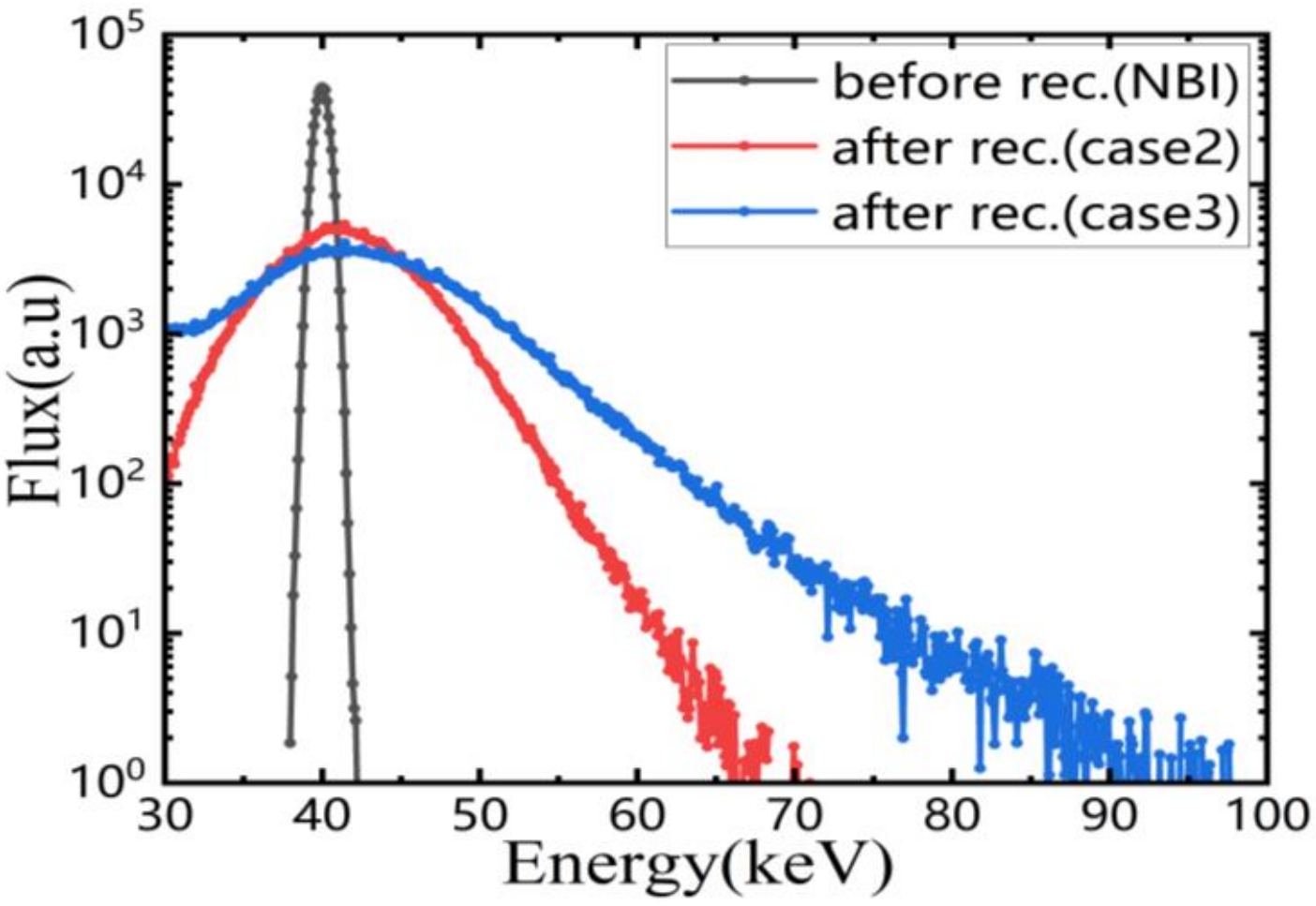

**FIG. 5.** Ion flux evolution for different reconnection scenarios in the simulation of the NBI discharge ($E_{inj}$=40keV). The black dotted line represents the initial energy distribution, while the red dotted and blue dotted lines represent the final energy distributions after reconnection in case 2 and case 3, respectively.

Another possible acceleration mechanism is the excitation of ion Bernstein waves (IBW) via nearly perpendicular beam injection[18]. However, in EXL-50U the neutral beam is injected nearly parallel to the magnetic field, making such excitation inefficient. Moreover, no IBW or ICRF band signals were detected by the high-frequency magnetic diagnostics during these discharges. Both theoretical simulation and experimental data indicate that wave-particle interactions are unlikely to be the acceleration mechanism for the high-energy particles observed in this experiment.

In conclusion, for the first time, we observed a stable and sustained acceleration of high-energy particles on the magnetic confinement device, without large-scale MHD activity or confinement degradation. Both simulation and experiment show that small-scale magnetic reconnection can accelerate pre-existing high-energy particles to higher energy. And during ECH, this acceleration is further enhanced by changing the local magnetic shear. Kinetic simulations incorporating the reconnection electric field qualitatively and semi-quantitatively reproduce the experimental observations. These findings reveal that small-scale magnetic reconnection serves as a potential means to stably accelerate non-thermal high-energy ions within magnetic confinement fusion devices. By augmenting both the energy and population of the high-energy tail, this mechanism holds promise for enhancing the gain of various fusion reactions, including D-T, D-$^{3}$He, and p-$^{11}$B.

## End Matter

*EXL-50U*

The EXL-50U spherical tokamak ($R_0$ = 0.7 m, $a$ = 0.4 m, $B_t$ =1.2 T) operated in hydrogen–boron plasmas with $I_p$ = 500 kA and $n_e$ =1 × $10^{19}$ $m^{-3}$. For the EXL-5U experiment presented in this paper, three heating schemes were employed during the plasma current flat-top phase: (i) pure Ohmic, (ii) 20-45 keV $H^0$ NBI heating, and (iii) simultaneous NBI plus 50 GHz EC resonance heating at $\rho_{ec} \approx$ 0.2–0.4.

*Experimental Setup and Diagnostics*

The experiments were conducted on the EXL-50U spherical torus. The $I_p$ and $V_{loop}$ were measured using Rogowski coils and flux loops, respectively. The electron density was monitored via a Thomson Scattering (TS) system, and the magnetic fluctuation signals were captured by Mirnov coils with a sampling rate of 500 kHz.

*Neutral Particle Analyzer (NPA) Measurement*

High-energy ion distributions were measured using a multi-channel *E//B* type Neutral Particle Analyzer (NPA). The NPA system utilizes a stripping cell to re-ionize neutral atoms escaping from the plasma, which are subsequently deflected by parallel electric and magnetic fields onto an array of 16 detectors. The analyzer was calibrated using a mono-energetic ion source prior to the campaign, ensuring an energy resolution of $\Delta E/E$ <10%[15]. The relative position between the NPA and the NBI is shown in Fig. 6.

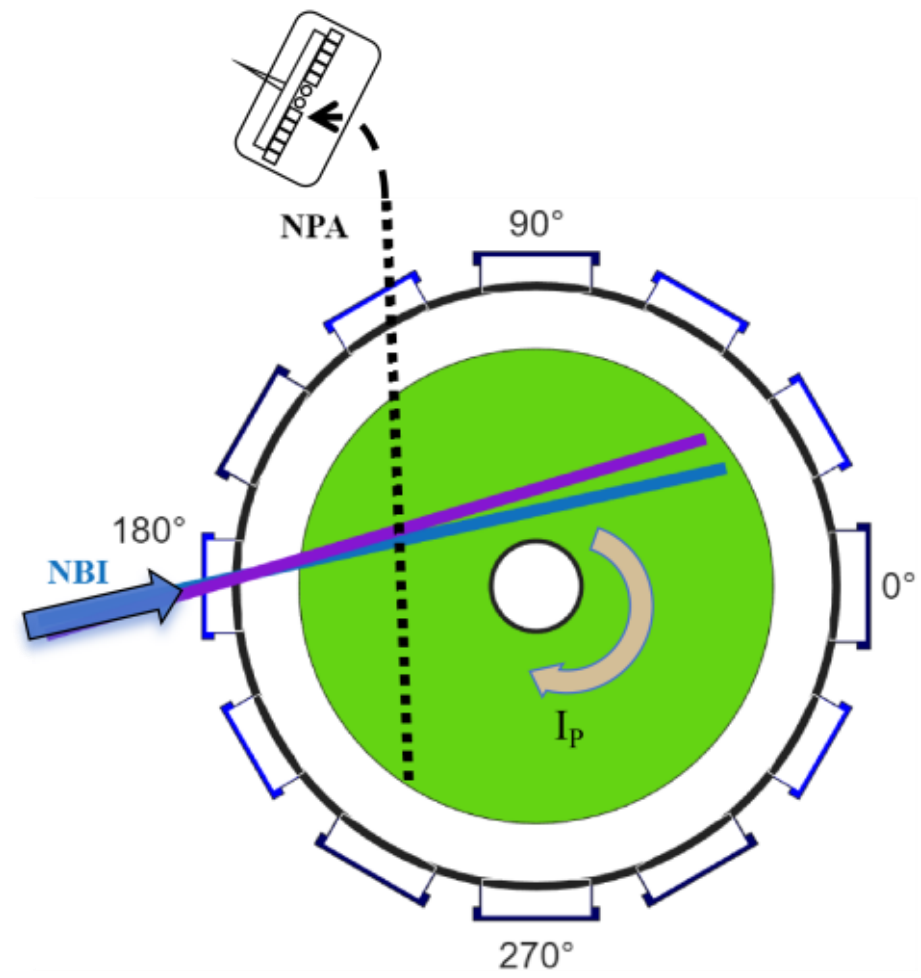

Figure 6. Relative position of the NPA with respect to the NBI.

*Kinetic Simulation Model*

Simulations are performed on the ***x-z*** plane, and the domain size is $L_x \times L_z$. The initial condition is a force-free current sheet with the magnetic field $B(z) = B_0 \tanh(z/\delta) e_x - \sqrt{B_0^2 \mathrm{sech}^2(z/\delta) + B_t^2 - B_0^2}\, e_y$ and a uniform density $n_0 = 5\times10^{18} m^{-3}$, where $B_0 = 0.21T$ and $B_t = 0.7T$. We use the real mass ratio $m_i/m_e = 1836$ and light speed $c = 3\times10^8 m/s$. We use the ion inertia length $d_i = c/\sqrt{n_0 e^2/\varepsilon_0 m_i} \approx 10cm$ and the inversed ion gyrofrequency $\Omega_i^{-1} = m_i/eB_t \approx 15ns$ to normalize the length and time during simulation. The initial temperature of ions and electrons are set to be $T_{i0} = T_{e0} = 1keV$. In case2 and 3, we add a high energy ion component with a uniform density $n_\mathrm{h} = 5\times10^{16} m^{-3}$ in the simulation domain. The kinetic energy of this component is $E_k = 0.5 m_i V_y^2 = 40keV$, and the half-width of the energy distribution of this component is $1keV$. To keep electric neutrality, an electron component with a uniform density $n_\mathrm{h}$ and temperature $T_{e0}$ is also added. Table 1 list the different parameters between the simulation cases.

Table 1. Some simulation parameters.

| | $L_x \times L_z$ | $\delta$ | Energetic ion |
|---|---|---|---|
| Case1 | $4d_i \times 2d_i$ | $0.05d_i$ | No |
| Case2 | $4d_i \times 0.5d_i$ | $0.05d_i$ | Yes |
| Case3 | $4d_i \times 0.5d_i$ | $0.025d_i$ | Yes |